\documentclass[12pt,a4,epsf]{article}
\global\arraycolsep=2pt 
\input{epsf}
\usepackage{graphicx}
\begin{document}
\makeatletter
\def\fmslash{\@ifnextchar[{\fmsl@sh}{\fmsl@sh[0mu]}}
\def\fmsl@sh[#1]#2{%
  \mathchoice
    {\@fmsl@sh\displaystyle{#1}{#2}}%
    {\@fmsl@sh\textstyle{#1}{#2}}%
    {\@fmsl@sh\scriptstyle{#1}{#2}}%
    {\@fmsl@sh\scriptscriptstyle{#1}{#2}}}
\def\@fmsl@sh#1#2#3{\m@th\ooalign{$\hfil#1\mkern#2/\hfil$\crcr$#1#3$}}
\makeatother
\thispagestyle{empty}
\begin{titlepage}
\begin{flushright}
hep-ph/0103333 \\
LMU 01/05 \\
\today
\end{flushright}

\vspace{0.3cm}
\boldmath
\begin{center}
\Large  {\bf  Electroweak D-Waves}
\end{center}
\unboldmath
\vspace{0.8cm}

\unboldmath
\vspace{0.8cm}
\begin{center}
  {\large Xavier Calmet}\\
  
\end{center}
\begin{center}
  and
  \end{center}
\begin{center}
{\large Harald Fritzsch}\\
 \end{center}
 \vspace{.3cm}
\begin{center}
{\sl Ludwig-Maximilians-University Munich, Sektion Physik}\\
{\sl Theresienstra{\ss}e 37, D-80333 Munich, Germany}
\end{center}
\vspace{\fill}
\begin{abstract}
\noindent
We consider phenomenological implications of a model recently proposed
for the electroweak interactions based on a $SU(2)_L$ confining
theory. We concentrate on the production of excited states of the
electroweak bosons at future colliders and we consider their contribution
to the reaction $W^+ + W^- \to W^+ + W^-$. We expect large deviations
from the standard model in the TeV region.
\end{abstract}
\end{titlepage}

\section{Introduction}
The aim of this work is to investigate the phenomenological
implications of a model recently proposed for the electroweak
interactions based on a $SU(2)_L$ confining theory
\cite{Calmet:2000th}. We shall concentrate on orbital and radial
excitations of the electroweak bosons. Our model makes use of the
confinement mechanism proposed in Ref.  \cite{'tHooft:1998pk}.  It was
emphasized in Ref.\cite{Calmet:2000vx} that models of a similar class imply
different search strategies for the Higgs boson than those usually
adopted when searching for the standard model, supersymmetric or
fermiophobic Higgs bosons.

In our model the left-handed particles appear as bound states of
fundamental, unobservable fermions $f_L$ and $q_L$ and a scalar $h$.
These particles transform as doublets under $SU(2)_L$. Besides this,
$q_L$ is a triplet under $SU(3)_c$. We can then identify the following
physical left-handed fermions, Higgs boson and electroweak bosons.
\begin{eqnarray} \label{def1}
\begin{array}{lllllll}
\mbox{neutrino:} &  \nu_L  & \rightarrow & \bar h l  & \\ 
 \mbox{electron:} &   e_L & \rightarrow& h l  & \\
 \mbox{up  type  quark:}&  u_L &\rightarrow  & \bar h q &  \\
 \mbox{down type quark:}&  d_L & \rightarrow & h q  & \\
 \mbox{Higgs particle:}& \phi & \rightarrow& \bar h h &
 (\mbox{$s$-wave}) \\
 \mbox{$W^3$ boson:} &   W^3 & \rightarrow& \bar h h &
 (\mbox{$p$-wave})  \\
 \mbox{$W^-$ boson:} &   W^- & \rightarrow& h h  &
 (\mbox{$p$-wave}) \\
 \mbox{$W^+$ boson:} &   W^+ & \rightarrow& (h h)^\dagger & (\mbox{$p$-wave}), 
\end{array}
\end{eqnarray}
assuming a $SU(2)_L$ confinement. The right-handed particles are those
of the standard model. In our approach the electroweak bosons appear
as excited states of the Higgs boson. It was shown in
Ref.\cite{Calmet:2000th} that the minimal sector of the model, i.e.,
the sector containing only the particles predicted by the standard
model, is identical to the standard model \cite{Glashow:1961tr} if one
chooses the unitary gauge, we call this property duality. This is done
by fixing the gauge and performing a $1/F$ expansion, where $F \cong
492$ GeV is the scale of the theory.  In this model new particles
corresponding to exotic particles like leptoquarks can be introduced.
But, they do not obey to this expansion, and the duality cannot be
applied to describe their properties. Forces between two fermions can
be very much different than those between a fermion and a scalar or
between two scalars. If leptoquarks do exist, their mass scale is
presumably very high.

Of particular interest are radially excited versions of the Higgs
boson $H^*$ and of the electroweak bosons $W^{3 *}$ and $W^{\pm*}$. As
described in \cite{Calmet:2000th}, the most promising candidates for
energies available at the LHC or at future linear colliders are the
excited states of the Higgs boson and of the electroweak bosons.
Especially the orbital excitation, i.e., the spin 2 $d$-waves
$D^3_{\mu \nu}$, $D^-_{\mu \nu}$ and $D^+_{\mu \nu}$, of the
electroweak bosons have a well defined $1/F$ expansion (we use the
unitary gauge: $h=(h_{(1)}+F,0)$:
\begin{eqnarray} \label{def3}
   D^3_{\mu \nu}&=& \frac{2}{g^2 F^2}
\left(   \left ( D_{\mu}h \right)^\dagger \left (D_\nu h\right) +
\left ( D_{\nu}h \right)^\dagger \left (D_\mu h\right)
\right)
   \approx b^3_\nu b^3_\mu + b^+_\nu b^-_\mu+  b^+_\mu b^-_\nu
    \\ \nonumber
   D^-_{\mu \nu}&=& \frac{ - \sqrt{2} }{g^2 F^2} \epsilon^{ij}
  \left(   \left ( D_{\mu}h \right)_i \left (D_\nu h\right)_j +
\left ( D_{\nu}h \right)_i \left (D_\mu h\right)_j
\right)
   \approx b^3_\mu b^-_\nu+ b^3_\nu b^-_\mu
  \\ \nonumber
   D^+_{\mu \nu}&=&\left(\frac{ - \sqrt{2} }{g^2 F^2} \epsilon^{ij}
  \left(   \left ( D_{\mu}h \right)_i \left (D_\nu h\right)_j +
\left ( D_{\nu}h \right)_i \left (D_\mu h\right)_j
\right) \right)^\dagger
   \approx  b^3_\mu b^+_\nu+ b^3_\nu b^+_\mu
\end{eqnarray}
where $D_{\mu}$ is the covariant derivative, $b_\mu^a$, $a=\{3,+,-\}$
are the gauge fields and $g$ the coupling constant corresponding to
the gauge group $SU(2)_L$. Although the masses and the couplings of
these electroweak $d$-waves to other particles are fixed by the
dynamics of the model, it is difficult to determine these parameters.
In analogy to Quantum Chromodynamics, it is expected that these
$d$-waves couple with a reasonable strength to the corresponding
$p$-waves, the electroweak bosons. In the following, we assume in
accordance with the duality property, that the $d$-waves only couple to
the electroweak bosons and not to the photon, Higgs boson or the
fermions.

\section{Production of the electroweak $d$-waves}

The cross-sections and decay width of $d$-waves predicted in a variety
of composite models were considered in Ref.\cite{Chiappetta:1987xc}.
Here we shall consider different effective couplings of our
electroweak $d$-waves that are more suitable for the model proposed in
\cite{Calmet:2000th}. If their masses are of the order of the scale of
the theory, they will be accessible at the LHC.  Of particular
interest is the neutral electroweak $d$-wave because it is expected to
couple to the $W^\pm$ electroweak bosons. This particle can thus be
produced by the fusion of two electroweak bosons at the LHC or at
linear colliders.

We  shall  use  the  formalism   developed  by  van  Dam  and  Veltman
\cite{vanDam:1970vg} for massive $d$-waves  to compute the decay width
of the $D^3_{\mu \nu}$ into $W^+ W^-$. We use the following relation:
\begin{eqnarray}
\sum^{5}_{i=1} e^i_{\mu \nu}(p) e^i_{\alpha \beta}(p) &=& \frac{1}{2}
\left (\delta_{\mu \alpha} \delta_{\nu \beta}
  + \delta_{\mu \beta}   \delta_{\nu \alpha}
  - \delta_{\mu \nu}\delta_{\alpha \beta}\right) \\ \nonumber
&&+\frac{1}{2} \left (\delta_{\mu \alpha} \frac{p_\nu p_\beta}{m_D^2}+
  \delta_{\nu \beta} \frac{p_\mu p_\alpha}{m_D^2}
  +  \delta_{\mu \beta} \frac{p_\nu p_\alpha}{m_D^2}
  + \delta_{\nu \alpha} \frac{p_\mu p_\beta}{m_D^2}
\right)
\\ \nonumber
&&+\frac{2}{3}
\left ( \frac{1}{2} \delta_{\mu \nu} -\frac{p_\mu p_\nu}{m_D^2}\right)
\left ( \frac{1}{2} \delta_{\alpha \beta} -\frac{p_\alpha p_\beta}{m_D^2}\right)
\end{eqnarray}
for the sum over the polarizations $e^i_{\mu \nu}$ of the $d$-wave.
In the notation of Ref.\cite{vanDam:1970vg} the sum over the
polarizations of the $W^\pm$ is given by
\begin{eqnarray}
  \sum^{3}_{i=1} e^i_{\mu}(p) e^i_{\nu}(p) &=& \delta_{\mu \nu}
  + \frac{p_\mu p_\nu}{m_W^2}
\end{eqnarray}
where $\delta_{\mu \nu}$ is the Euclidean metric. 
Averaging over the polarizations of the $d$-wave, we obtain
\begin{eqnarray}
  \Gamma(D^3\rightarrow W^+ W^-)&=&
  \frac{g^2_D}{1920 m_D \pi} \left(x_W-4\right)^2
  \sqrt{\left(1-\frac{4}{x_W}\right)}
\end{eqnarray}
with $x_W=(m_D/m_W)^2$, where $m_D$ is the mass of the $d$-wave and
$g_D$ is a dimensionfull coupling constant with $\mbox{dim}[
g_D]=\mbox{GeV}$. A dimensionless coupling constant is obtained by a
redefinition of the coupling constant $g_D \rightarrow m_D \bar{g}_D$.
We shall discuss plausible numerical inputs in the next section.
Assuming that the $Z$ boson couples with the same strength to the
$d$-wave as the $W$-bosons, we can approximate the decay width into
$Z$ bosons in the following way
\begin{eqnarray}
 \Gamma(D^3\rightarrow Z Z)&=&
  \frac{g^2_D}{3840 m_D \pi} \left(x_Z-4\right)^2
  \sqrt{\left(1-\frac{4}{x_Z}\right)} \\ \nonumber
   &\approx& \frac{1}{2} \Gamma(D^3\rightarrow W^+ W^-) 
\end{eqnarray}
with $x_Z=(m_D/m_Z)^2$. The Breit-Wigner resonance cross section for
the reaction $W^+ + W^- \rightarrow D^3$ thus reads (see e.g.
\cite{Donoghue:1992dd})
\begin{eqnarray}
\sigma^{(res)}_{W^+ + W^-\rightarrow D^3}&=& \frac{10 \pi}{ q^2} \frac{m_D^2 \Gamma^{\mbox{(tot)}}_D \Gamma(D^3\rightarrow W^+ W^-)}{\left(m_D^2-s\right)^2+m_D^2 {\Gamma^{\mbox{(tot)}}_D}^2}
\end{eqnarray}
where $q^2=(s-4 m_W^2)/4$ and $\Gamma^{(\mbox{tot})}_D\approx 3/2
\Gamma(D^3\rightarrow W^+ W^-)$ is the total decay width of the
neutral $d$-wave. Due to the background, the $W$ bosons might be
difficult to observe. But, if the electroweak $d$-waves states are
produced we expect an excess of $Z$ bosons compared to the standard
model expectation. Note that the $Z$ bosons are easier to observe.

As we shall see in the next section, the neutral $d$-waves give a sizable
contribution to the reaction $W^+ + W^- \to W^+ + W^-$.

\section{The reaction $W^+ + W^- \to W^+ + W^-$}

A considerable attention has been paid to the scattering of
electroweak bosons since this represents a stringent test of the gauge
structure of the standard model. In particular the reaction $W^+ + W^-
\to W^+ + W^-$ is known to be of prime interest. If the Higgs boson is
heavier than 1 TeV, the electroweak bosons will start to interact
strongly \cite{Dicus:1973vj}. This reaction has been studied in the
framework of the standard model in Ref.  \cite{Duncan:1986vj} and the
one loop corrections were considered in \cite{Denner:1998kq} and are
known to be sizable. For the sake of this paper the tree level
diagrams are sufficient to show that the contribution of the neutral
electroweak $d$-wave will have a considerable impact to that reaction
and cannot be overlooked in forthcoming experiments. As described in
\cite{Gunion:1989we} (see also Ref.  \cite{Duncan:1986vj}) the $W$'s
emitted by the beam particles are dominantly longitudinally polarized
if the following relations are fulfilled: $m_W^2 \ll m^2_{WW} \ll s$
at an $e^+$ $e^-$ collider, and $m_W^2 \ll m^2_{WW} \ll s_{q \bar q}
\ll s$ at a hadron collider, and we shall only consider the
especially interesting reaction $W^+_L + W^-_L \to W^+_L + W^-_L$ as
described in \cite{Duncan:1986vj}.

\begin{figure}
  \begin{minipage}[t]{0.32\linewidth}\centering
    \includegraphics[width=\linewidth]{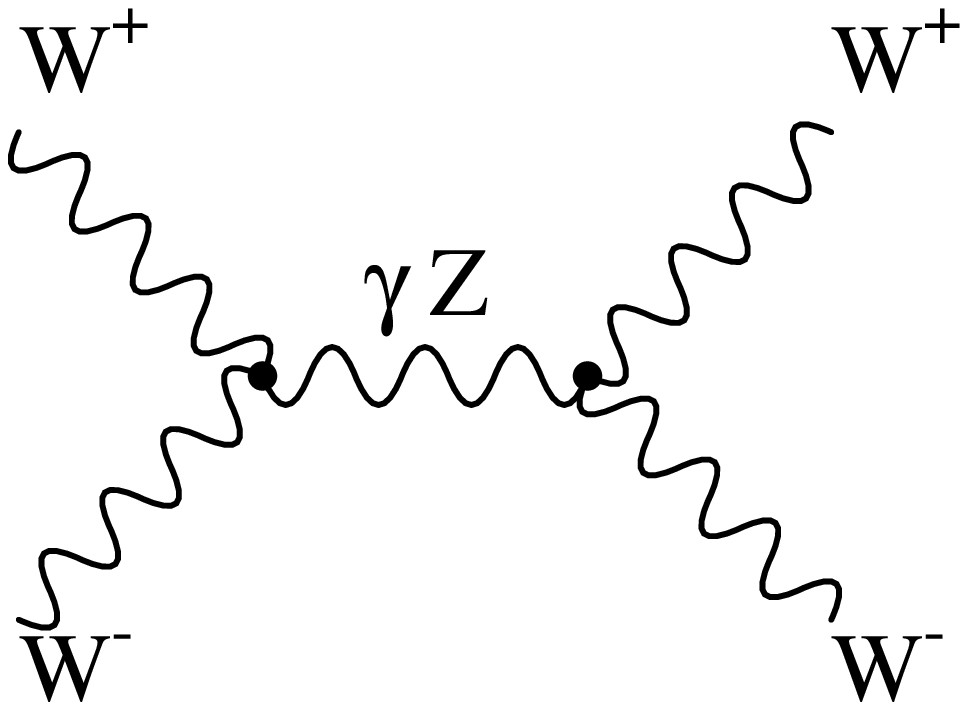}
    \begin{minipage}{1\linewidth}
      \caption{photon and Z boson in the s channel \label{fig1}}
    \end{minipage}
  \end{minipage}
  \begin{minipage}[t]{0.32\linewidth}\centering
    \includegraphics[width=\linewidth]{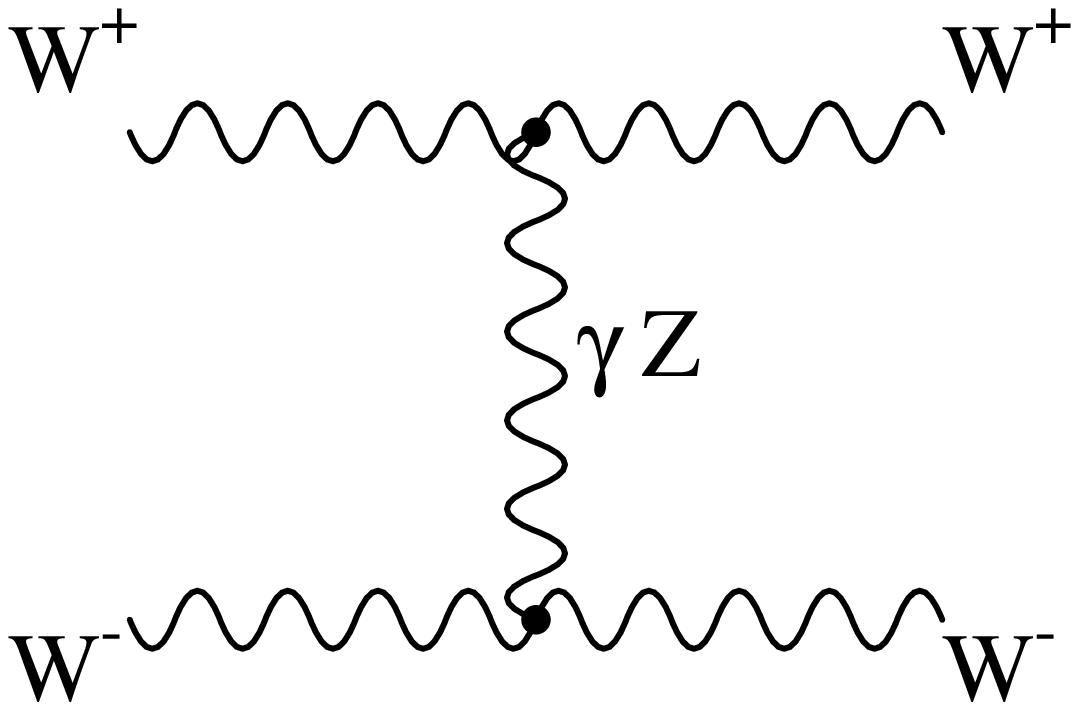}
    \begin{minipage}{1\linewidth}
      \caption{photon and Z boson in the t channel\label{fig2}}
    \end{minipage}
    \end{minipage}
  \begin{minipage}[t]{0.32\linewidth}\centering
    \includegraphics[width=\linewidth]{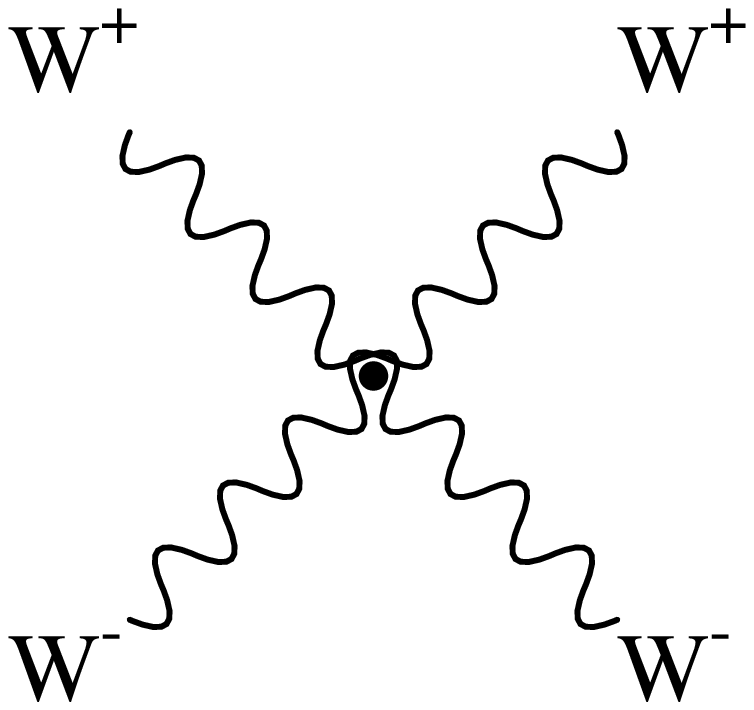}
    \begin{minipage}{1\linewidth}
      \caption{four W vertex \label{fig3}}
    \end{minipage}
  \end{minipage}
  \end{figure}

\begin{figure}
  \begin{minipage}[t]{0.45\linewidth}\centering
    \includegraphics[width=\linewidth]{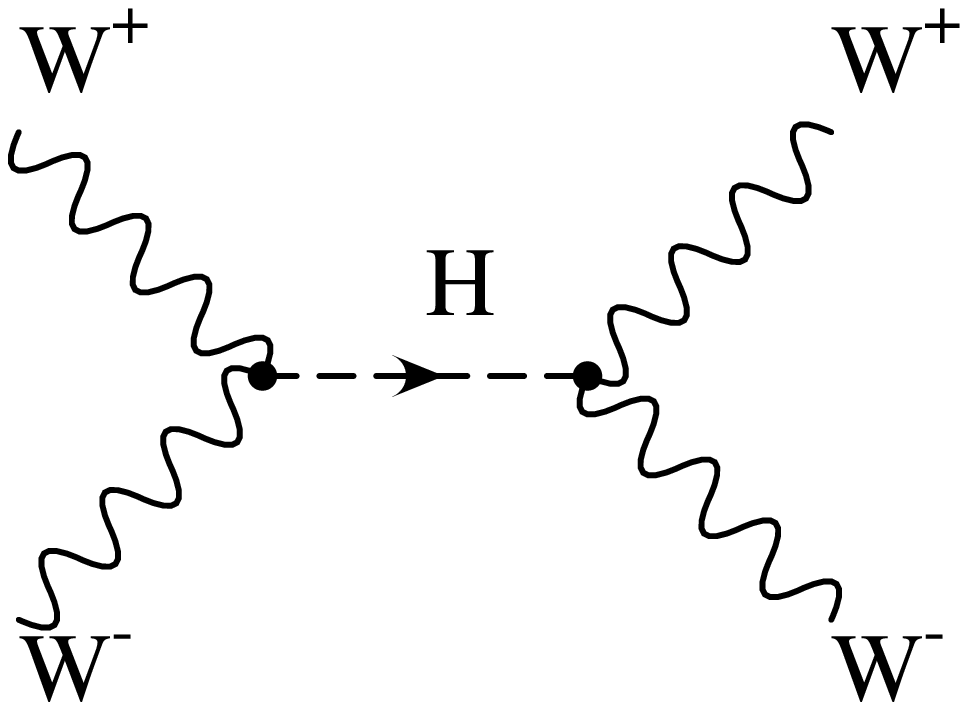}
    \begin{minipage}{0.9\linewidth}
      \caption{Higgs boson in the s channel\label{fig4}}
    \end{minipage}
    \end{minipage}
     \begin{minipage}[t]{0.45\linewidth}\centering
    \includegraphics[width=\linewidth]{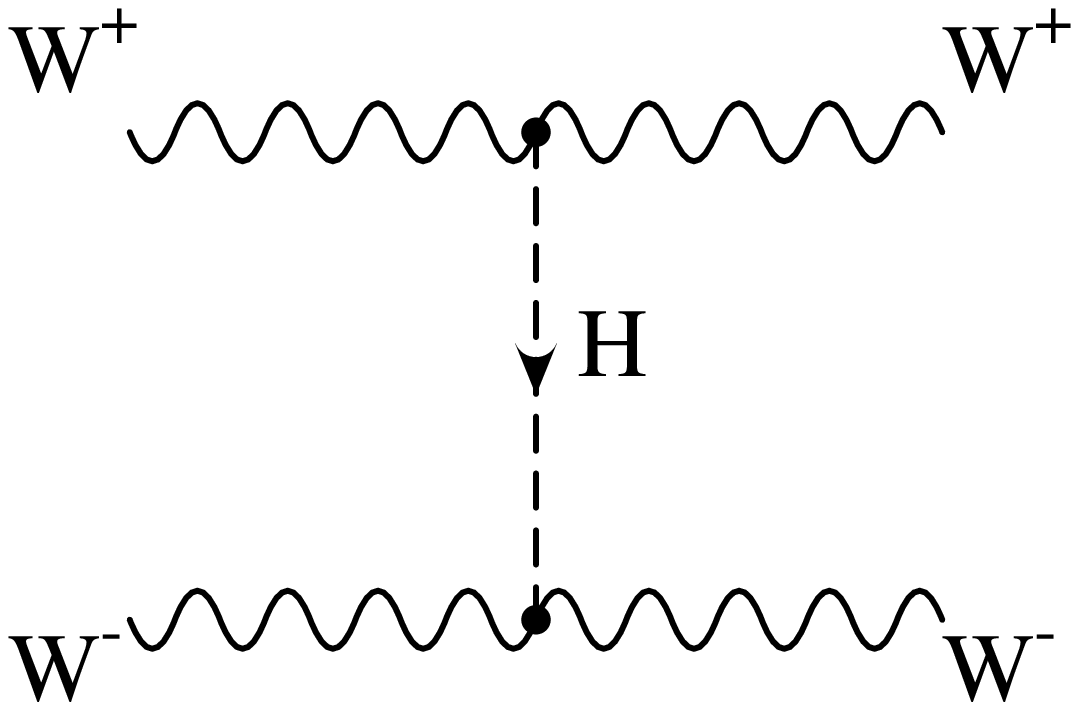}
    \begin{minipage}{0.9\linewidth}
      \caption{Higgs boson in the t channel\label{fig5}}
    \end{minipage}
  \end{minipage}
  \end{figure}

In the standard model, this reaction is a test of the gauge structure
of the theory \cite{LlewellynSmith:1973ey}. The Feynman graphs
contributing in the standard model to this reaction can be found in
figures \ref{fig1}, \ref{fig2}, \ref{fig3}, \ref{fig4} and
\ref{fig5}.
The amplitudes corresponding to these graphs are \cite{Duncan:1986vj}
\begin{eqnarray}
A_{s \gamma} &=& -\frac{1}{16}i g^2 x s^2 \beta^2 (3-\beta^2)^2 \cos{\theta},
\\ \nonumber 
 A_{s Z}&=& -\frac{1}{16}i g^2 (1-x) \frac{s^3}{s-\xi_Z} \beta^2
  (3-\beta^2)^2 \cos{\theta},
 \\ \nonumber  
  A_{t \gamma} &=& -\frac{1}{32 }i g^2 x \frac{s^3}{t}
                   [\beta^2(4- 2 \beta^2+\beta^4)   
                   +\beta^2(4- 10 \beta^2+\beta^4) \cos{\theta} \\
                    \nonumber &&
                   +(2-11 \beta^2+10\beta^4) \cos^2{\theta}
                   +\beta^2 \cos^3{\theta}],
\\ \nonumber 
A_{t Z} &=& -\frac{1}{32 }i
                 g^2 (1-x) \frac{s^3}{t - \xi_Z}
                   [\beta^2(4- 2 \beta^2+\beta^4)   
                   +\beta^2(4- 10 \beta^2+\beta^4) \cos{\theta}
                   \\
                    \nonumber &&
                   +(2-11 \beta^2+10\beta^4) \cos^2{\theta}
                   +\beta^2 \cos^3{\theta}],
\\ \nonumber 
   A_4&=& -\frac{1}{16} i g^2 s^2
   (1 + 2 \beta^2 -6 \beta^2 \cos{\theta}- \cos^2{\theta}),
\\ \nonumber 
     A_{s H}&=&-\frac{1}{16} i g^2 s^2
     \frac{(1+\beta^2)^2}{s-\xi_H+i\gamma_H}, 
\\ \nonumber 
     A_{t H}&=&-\frac{1}{16} i g^2 s^2 \frac{(\beta^2-\cos{\theta})^2}
     {t-\xi_H+i\gamma_H},
\end{eqnarray}
where $x=\sin^2{\theta_W}$, $\xi_Z=(1-x)^{-1}=m_Z^2/m^2_W$,
$\xi_H=m_H^2/m^2_W$, $\gamma_H=m_H \Gamma_H/m_W^2$ and
$\beta=\sqrt{1-4/s}$. The variables $s$ and $t$ are scaled with
respect to $m_W^2$. The scattering angle is $\theta$, $t=-1/2 s
\beta^2(1-\cos{\theta})$. These notations are the same as those
introduced in Ref.\cite{Duncan:1986vj}.  The standard model amplitude
is thus
\begin{eqnarray}
  A^{sum}_{SM}&=&A_{s \gamma}+A_{s Z}+A_{t \gamma}+A_{t Z}+A_{4}
  +A_{s H}+A_{t H}.
\end{eqnarray}
In the high energy limit, one observes the cancellation of the
leading powers in $s$ and finds \cite{Duncan:1986vj}
\begin{eqnarray}
  A^{sum}_{SM} \approx \frac{1}{2} i g^2 \left[ \xi_Z \left(1+ \frac{s}{t}
      + \frac{t}{s} \right) + \xi_H - i \gamma_H \right]
\end{eqnarray}
for the sum of these amplitudes.
The cross section with the angular cut $-z_0< \cos\theta<z_0$ is then
\begin{eqnarray}
\sigma&=& \frac{1}{16 \pi s^2 \beta^2} \int^{t_+}_{t_-} |A^{sum}|^2 \mbox{d}t
\end{eqnarray}
in dimensionless units, $t_\pm=(2-s/2)(1\mp z_0)$.

The excitations of the Higgs and electroweak bosons also contribute
via the $s$ and $t$ channel.  The amplitudes corresponding to the
contribution of a radially excited Higgs boson ($H^*$) of mass
$m_{H^*}$ and decay width $\Gamma_{H^*}$ to this reaction are
\begin{eqnarray}
     A_{s H^*}&=&-\frac{1}{16} i g^2_{H^*} s^2
     \frac{(1+\beta^2)^2}{s-\xi_{H^*}+i\gamma_{H^*}}, 
\\ \nonumber 
     A_{t H^*}&=&-\frac{1}{16} i g^2_{H^*} s^2 \frac{(\beta^2-\cos{\theta})^2}
     {t-\xi_{H^*}+i\gamma_{H^*}},
\end{eqnarray}
where $\xi_{H^*}=m_{H^*}^2/m^2_W$, $\gamma_{H^*}=m_{H^*} \Gamma_{H^*}/m_W^2$
and $g_{H^{*}}$ is the strength of the coupling between two $W$ bosons
and the $H^{*}$ scalar particle.

We shall now consider the contribution of the
radially $(W^{3 *})$ and orbitally $(D^{\mu \nu})$ excited neutral $Z$
boson. The amplitudes for the $W^{3 *}$ can be at once deduced from
those of the standard model contribution of the $Z$ boson
\begin{eqnarray}
  A_{s W^{3 *}}&=& -\frac{1}{16}i g^2_{W^{3 *}}
  \frac{s^3}
    {s-\xi_{W^{3 *}}+ i \gamma_{W^{3 *}}} \beta^2
  (3-\beta^2)^2 \cos{\theta},
\\
  \nonumber 
 A_{t W^{3 *}} &=& -\frac{1}{32 }i
                 g^2_{W^{3 *}} \frac{s^3}
                 {t - \xi_{W^{3 *}}+i\gamma_{W^{3 *}}}
                   [\beta^2(4- 2 \beta^2+\beta^4)
                   \\
                    \nonumber &&
                   +\beta^2(4- 10 \beta^2+\beta^4) \cos{\theta}
                   +(2-11 \beta^2+10\beta^4) \cos^2{\theta}
                   +\beta^2 \cos^3{\theta}],
\end{eqnarray}
where $\xi_{W^{3 *}}=m_{W^{3 *}}^2/m^2_W$, $\gamma_{W^{3 *}}=m_{W^{3 *}}
\Gamma_{W^{3 *}}/m_W^2$ and $g_{W^{3 *}}$ is the strength of the coupling
between two $W$ bosons and the $W^{3 *}$ boson.

The orbitally excited $Z$ boson $(D^{\mu \nu})$ is a $d$-wave, and its
propagation is thus described by a propagator corresponding to a
massive spin 2 particle. The propagator of a massive spin two particle
is as follows (see Ref. \cite{vanDam:1970vg}):
\begin{eqnarray}
  \Gamma_{\mu \nu \rho \sigma} &=& \frac{1}{p^2-m_D^2}
   \frac{1}{2} ( g_{\mu \rho} g_{\nu \sigma} +
                 g_{\mu \sigma} g_{\nu \rho}- \frac{2}{3}
                 g_{\mu \nu} g_{\rho \sigma})
\end{eqnarray}
and we assume that the vertex $W^{+ \mu} W^{- \nu} D_{\mu \nu}$ is of
the form $i g_D$.  We obtain the following amplitudes for
the $s$ and $t$ channel exchange
\begin{eqnarray}
A_{s D}&=&\frac{-1}{48} i g_D^2 \frac{m_D^2}{m_W^2} 
\frac{s^2}{s-\xi_D+i \gamma_D}
\left(2\beta^4+3 \cos^2{\theta} - 2 \beta^2-1\right)
\end{eqnarray}
\begin{eqnarray}
\! \! \! \! \! \! \! \! \!
A_{t D}&=& \frac{-1}{96} i g_D^2  \frac{m_D^2}{m_W^2} 
\frac{s^2}{t-\xi_D+i \gamma_D}
\left(
  4\beta^4+6\beta^2+3+10\beta^2\cos{\theta}+1\cos{\theta}^2 \right)
\end{eqnarray}

Since there is a pole in the $t$ channel whose origin is the photon
exchange, one has to impose cuts on the cross sections. For the
numerical evaluation of the cross section, we impose a cut of
$10^\circ$, which is the cut chosen in Ref.  \cite{Denner:1998kq}. The
spin of the particle can be determined from the angular distribution
of the cross section. We have neglected the decay width of the $Z$
boson and that of the Higgs boson since we assume that the energy of the
process is such that no $Z$ boson or Higgs resonance appear. For
numerical estimates, we took $m_H=100$ GeV.

We have considered only the reaction involving longitudinally polarized 
$W$. The amplitudes for different polarizations for the standard model can 
be found in the literature \cite{Duncan:1986vj}. The amplitudes for a $H^*$ 
or a $W^{3*}$ can be deduced from the standard model calculations by 
replacing the masses, the decay widths and the coupling constants. Those 
for the neutral $d$-wave can be easily calculated using

\begin{eqnarray}
{\cal A}^{p_1 p_2 p_3 p_4}_s&=&-i g_D^2 \frac{m_D^2}{m_W^2} 
  \frac{1}{s-\xi_D+i\gamma_D} \\ \nonumber &&
   \frac{1}{2}\epsilon^\mu(p_1) \epsilon^\nu(p_2)
   ( g_{\mu \rho} g_{\nu \sigma} +
                 g_{\mu \sigma} g_{\nu \rho}- \frac{2}{3}
                 g_{\mu \nu} g_{\rho \sigma})
                 \epsilon^{* \rho}(p_3) \epsilon^{* \sigma}(p_4)
\end{eqnarray}
and
\begin{eqnarray}
{\cal A}^{p_1 p_2 p_3 p_4}_t&=&-i g_D^2  \frac{m_D^2}{m_W^2} 
  \frac{1}{t-\xi_D+i\gamma_D} \\ \nonumber &&
   \frac{1}{2}\epsilon^\mu(p_1) \epsilon^\rho(p_2)
   ( g_{\mu \rho} g_{\nu \sigma} +
                 g_{\mu \sigma} g_{\nu \rho}- \frac{2}{3}
                 g_{\mu \nu} g_{\rho \sigma})
                 \epsilon^{* \nu}(p_3) \epsilon^{* \sigma}(p_4)
\end{eqnarray}
where $p_i$ stands for the polarization and also using the following
relations
\begin{eqnarray}
  &\epsilon^\mu_1(0)=(-p,0,0,E)/m_W
  &\epsilon^\mu_1(\pm)=(0,-1,\pm i,0)/\sqrt{2}           \\
  \nonumber
   &\epsilon^\mu_2(0)=(-p,0,0,-E)/m_W
  &\epsilon^\mu_2(\pm)=(0,1,\pm i,0)/\sqrt{2} \\
  \nonumber
   &\epsilon^{*\mu}_{3}(0)=(p,-E \sin{\theta},0,-E\cos{\theta} )/m_W
  &\epsilon^{*\mu}_{3}(\pm)=(0,-\cos{\theta},\mp i,\sin{\theta})/\sqrt{2}           \\
  \nonumber
   &\epsilon^{*\mu}_4(0)=(p,E \sin{\theta},0,E\cos{\theta} )/m_W
  &\epsilon^{*\mu}_4(\pm)=(0,\cos{\theta},\mp i,-\sin{\theta})/\sqrt{2}
\end{eqnarray}
valid in the center of mass system where $E$ is the energy of the $W$
bosons, $p=\sqrt{E^2-m_W^2}$ is their momentum and $\theta$ is the
scattering angle.

\section{Discussion}

\begin{figure}
\begin{center}
\leavevmode
\epsfxsize=8cm
\epsffile{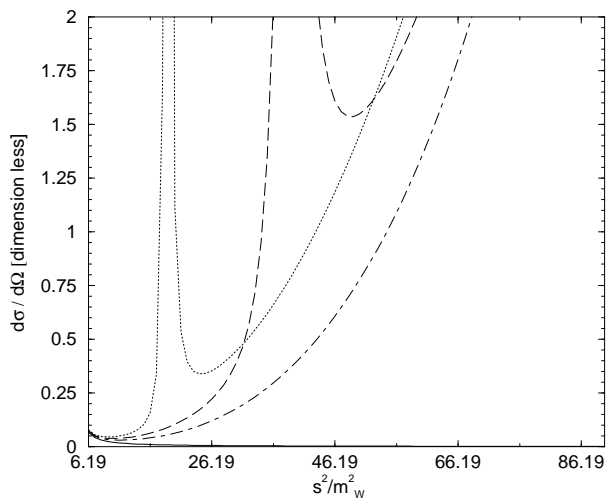}
\caption{Dimensionless cross section of the reaction $W_L^+ W_L^- \to W_L^+
  + W_L^-$ including the d-wave. The solid line is the standard model
  cross section, the dotted line corresponds to a d-wave of mass 350
  GeV, with $\Gamma=4.38$ GeV and $\bar{g}_{W^{3 *}}=0.8 g$, the long dashed
  line to a d-wave of mass 500 GeV, with $\Gamma=27.49$ GeV and $\bar{g}_{W^{3
      *}}=0.7 g$ and the dot-dashed line to a d-wave of mass 800 GeV,
  with $\Gamma=251.03$ GeV and $\bar{g}_{W^{3 *}}=0.6 g$.}
\label{dwave}
\end{center}
\end{figure}

\begin{figure}
\begin{center}
\leavevmode
\epsfxsize=8cm
\epsffile{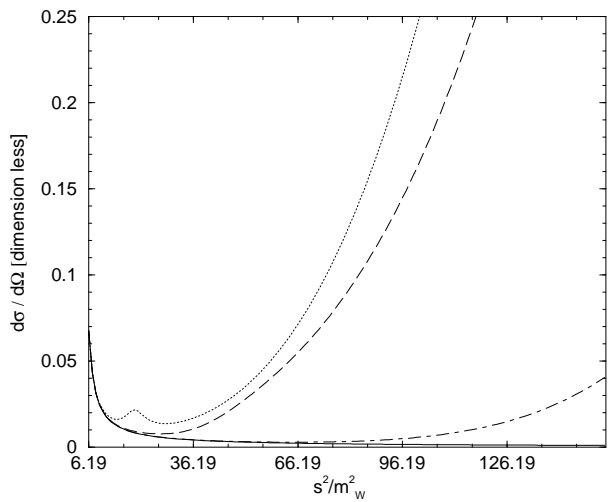}
\caption{Dimensionless cross section of the reaction
  $W_L^+ W_L^- \to W_L^+ + W_L^-$
  including the $W^{3*}$ boson. The solid line is the standard model
  cross section, the dotted line corresponds to a $W^{3*}$ boson of
  mass 350 GeV, with $\Gamma=66.2$ GeV and $\bar{g}_{W^{3 *}}=0.8
  \sin^2{\theta_W} g$, the long dashed line to a $W^{3*}$ boson of mass
  500 GeV, with $\Gamma=266.8$ GeV and $\bar{g}_{W^{3 *}}=0.7 \sin^2{\theta_W}
  g$ and the dot-dashed line to a $W^{3*}$ boson of mass 800 GeV, with
  $\Gamma=1795.5$ GeV and $\bar{g}_{W^{3 *}}=0.6 \sin^2{\theta_W} g$.}
\label{zprime}
\end{center}
\end{figure}

\begin{figure}
\begin{center}
\leavevmode
\epsfxsize=8cm
\epsffile{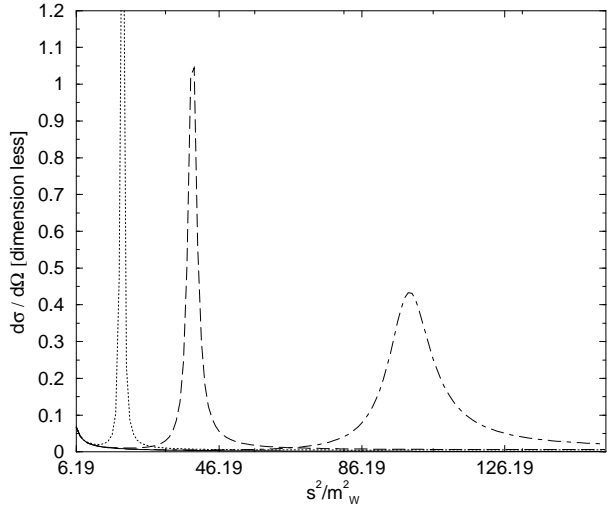}
\caption{Dimensionless cross section of the reaction
  $W_L^+ W_L^- \to W_L^+ + W_L^-$
  including the $H^{*}$ boson. The solid line is the standard model
  cross section, the dotted line corresponds to a $H^{*}$ boson of
  mass 350 GeV, with $\Gamma=6.72$ GeV and $\bar{g}_{H^{*}}=0.8 g$, the long
  dashed line to a $H^{*}$ boson of mass 500 GeV, with $\Gamma=17.6$
  GeV and $\bar{g}_{H^{*}}=0.7 g$ and the dot-dashed line to a $W^{3*}$
  boson of mass 800 GeV, with $\Gamma=58.25$ GeV and $\bar{g}_{H^{*}}=0.6 g$.}
\label{Hstar}
\end{center}
\end{figure}

The differential decay widths for the reaction $W_L^+ W_L^- \to W_L^+
+ W_L^-$ can be found in figure \ref{dwave} for the reaction involving
the neutral $d$-wave, figure \ref{zprime} for that involving the
$W^{3*}$ spin 1 boson and figure \ref{Hstar} for that involving the
$H^{*}$ scalar. The particles $W^{3*}$ and $H^{*}$ are assumed to
couple, in a first approximation, only to the $W$'s. This allows to
compute their decay rates using standard model formulas. As mentioned
previously it is not an easy task to predict the mass spectrum of the
model, thus we assumed, for numerical illustration, three different
masses: 350 GeV, 500 GeV and 800 GeV. The coupling constants are
assumed to sizable (see figures).  If the cross sections are
extrapolated to very high energies, the unitarity is violated.
However, as expected in any substructure models, it will be restored by
bound states effects.

\begin{figure}
\begin{center}
  \leavevmode \epsfxsize=8cm \epsffile{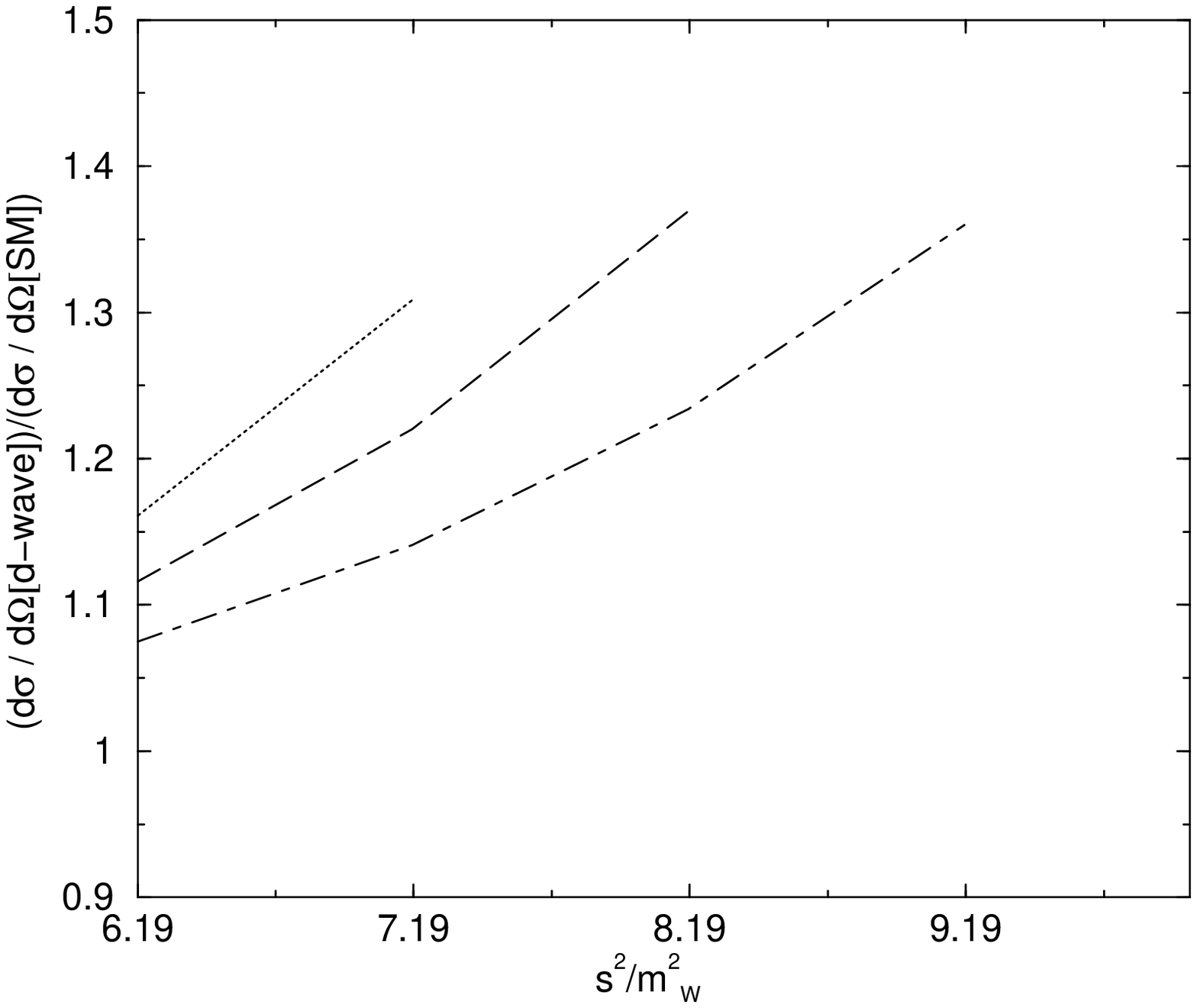}
\caption{Ratio of the cross-section for the of the reaction involving the $d$-wave to the standard model cross-section for different values of the $d$-wave mass and different coupling constant. The dotted line corresponds to a d-wave of mass 350 GeV, with $\Gamma=4.38$ GeV and $\bar{g}_{W^{3 *}}=0.8 g$, the long dashed  line to a d-wave of mass 500 GeV, with $\Gamma=27.49$ GeV and
  $\bar{g}_{W^{3 *}}=0.7 g$ and the dot-dashed line to a d-wave of
  mass 800 GeV, with $\Gamma=251.03$ GeV and $\bar{g}_{W^{3 *}}=0.6
  g$.}
\label{div}
\end{center}
\end{figure}

It is very instructive to plot the ratio of the differential cross
section involving new physics to the standard model differential cross
section. We have done so for the neutral $d$-wave (Ref. \ref{div}). It
is obvious from this picture that any deviation from the standard
model, even at high energy will manifest it-self already in a
deviation from one for that ratio. Already at an energy which is low
compared to the mass of the new particle, i.e. well bellow the
resonance, one observes a deviation from unity.

Nevertheless the calculation of the full reaction e.g. $e^+ e^- \to
W^+ W^- \nu \bar \nu$ involves the convolution of the cross section of
the reaction $W^+ W^- \to W^+ + W^-$ with functions describing the
radiative emission of the $W$'s from the fermions. When this integral
is performed some sensitivity is lost. Nevertheless the effects are
expected to be so large that they cannot be overlooked.  The reaction
will allow to test a mass range of a few TeV's so that even if the new
particles are too massive to be produced on-shell, their effects will
be noticeable at future colliders.

\section{Conclusions}

We have discussed the production of a neutral $d$-wave $D^3$ at the
LHC or at a linear collider. If the mass of this particle is of the
order of the scale of the theory, i.e. 300 GeV, it can be produced at
these colliders.  We have also shown that this particle as well as
radial excitations of the Higgs boson and $Z$ boson would spoil the
cancellation of the leading powers in $s$ of in the reaction $W^+_L
+W^-_L \to W^+_L +W^-_L$, thus any new particle contributing to that
reaction will have a large impact already at energies well below the
mass of this new particle. This reaction is thus not only of prime
interest if the Higgs boson is heavy but should also be studied if the
Higgs boson was light.

\section*{Acknowledgements}
We should like to thank P. Bambade, G. L. Kane, A. Leike, Z. Xing, V.
I. Zakharov and P. Zerwas for useful discussions.

\end{document}